\documentclass[12pt]{article}
\usepackage{graphicx}
\begin{document}
\centerline {\large \bf Sociophysics Simulations}

\bigskip
\centerline{Dietrich Stauffer, Institute for Theoretical Physics} 

\centerline{Cologne University, D-50923 K\"oln, Euroland}

\bigskip
Abstract: Check if we need one
\bigskip

\centerline {Introduction:}

Physicists since a long time have tried to apply their skills to fields outside
physics, with more or less success. Econophysics is a big fashion at present,
though partly based on the wrong belief that economists did not make empirical
observations of financial markets or made no Monte Carlo simulations where
rational judgement is replaced by random decisions. Also  Frederick Soddy 
ventured 
into economics after his 1921 Nobel prize in chemistry. Quantum mechanics
co-inventor Erwin Schr\"odinger more than half a century ago wrote a book
asking ``What is life ?'', long before today's interests in biophysics and
bioinformatics. Sociophysics has been around for at least three decades, with
or without that name \cite{schelling,weidlich,callen,galam}. The present review
summarizes some of the more recent simulations in sociophysics and is clearly
biased by the personal preferences and experience of its author. In June 2002,
I visited the first conference devoted only to Sociophysics, organized by
F. Schweitzer and K.G. Treutsch, {\tt www.ais.fhg.de/~frank/}. We ignore 
here car traffic simulations \cite{chowdhury,helbing}, scale-free networks 
\cite{albert}, social percolation \cite{weisbuchsolomon} and active Brownian 
particles \cite{schweitzer} since they were
reviewed recently. Instead we look at the models of Bonabeau et al 
\cite{bonabeau}, Sznajd \cite{sznajd} and similar consensus models 
\cite{deffuant,hegselmann}. We concentrate on simple models which take about
one page of Fortran program, available from {\tt stauffer@thp.uni-koeln.de}. We
thus update similar summaries published before \cite{moss,coniglio}.

\centerline{Hierarchies:}
How come someone is born into nobility, and others are members of the 
proletariat.
Some scientists are given tenure, and others have to leave academia
The elites of all countries and all times has always some explanations, like
the Grace of God having them put into the upper levels of society. Statistical
Physicists, of course, assume these hierarchies to arise from randomness.
(The illusion that everything is random is a professional disease ''morbus
Boltzmann'' among these physicists, just as silicosis = black lung affects
mine corkers.)

If nobility is connected with ownership of the land, then it cannot develop
easily in a nomadic society, while sedentary societies may have ground
property. The peasants then can become slaves of the nobility owning the piece
of territory on which the peasants work. Thus sedentary societies with 
agricultural fields may develop stronger hierarchies than nomadic tribes
with just a few light goods to be carried  around. The Hollywood movie 
``Dances with Wolves'' is an example how the industrialized world may imagine
beautiful nomadic paradise to have been. 

The problem therefore is to develop a model giving rise to strong hierarchies
at high population densities and weak hierarchies at low population densities.
Such a model, with a sharp first-order phase transition at some critical
density, was proposed by Bonabeau et al \cite{bonabeau} and followed-up by
others \cite{sousa,bonaparis}. People diffuse on a square lattice to nearest
neighbours (Margittai Neumann neighbourhood, not Moore neighbourhood). When
one person tries to move to a place already occupied by another
person, the two have a fight, the winner takes the contested 
place, and the loser moves to (or stays at) the other place. Initially, the
probability to win or lose is 50 percent. But after some time, memories of 
past fights and their outcomes accumulate and are stored in the history $h(i)$
for each person $i$, where $h$ is the number of victories minus the number
of defeats of this person. After each iteration, $h$ is diminished by, say,
10 percent of its current value to take into account that memories fade
away. 

\begin{figure}[hbt]
\begin{center}
\includegraphics[angle=-90,scale=0.5]{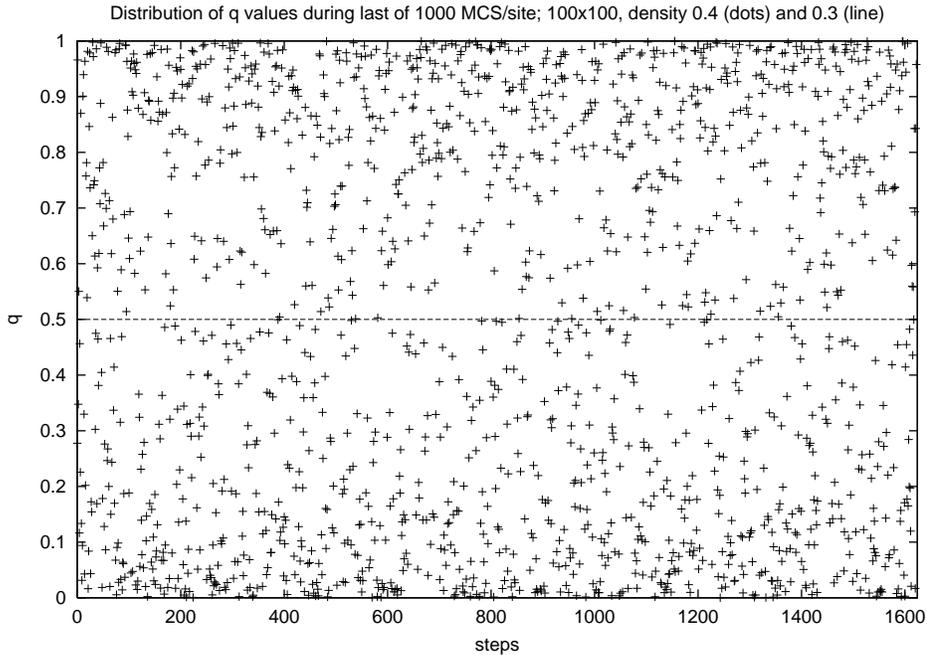}
\end{center}
\caption{Inequalities in the Bonabeau model, for density 0.4. The straight
line symbolizes the absence of hierarchies (all $q = 1/2$) for the lower
density 0.3. Only at the higher density do the probabilities $q$ to win a
fight concentrate on new very low and very high values.
}
\end{figure}

The probability $q$ for $i$ to win against $k$ is assumed as
$$q = 1/(1 + \exp(\sigma [h(k)-h(i)]) \eqno (1)$$
like a Fermi function: More victories in the past mean a higher chance to
win now. Here $\sigma$, the inverse width of the transition at Fermi level,
is the rms fluctuation in the winning probabilities:
$$\sigma = (<q^2> - <q>^2)^{1/2} \eqno (2)  $$
and thus is zero if everybody is equal ($q = 1/2$), and is $1/\sqrt 6$ if the 
values for $q$ are distributed homogeneously between zero and unity. Thus
$\sigma$ measures the inequality in this society. And this inequality 
re-enters into the probabilities to win and lose, thus enhancing existing
inequalities. (For the first ten Monte Carlo steps per site, $\sigma = 1$
to allow a build-up of hierarchies.)

The simulations show either a $\sigma$ going after some initial positive
values rapidly to zero with time (density below 32 percent), or a $\sigma$
staying at or above 0.25 for density above 32 percent. This first-order 
phase transition actually has a complicated history \cite{bonabeau,sousa,
bonaparis}, and only the present status is summarized here. Fig.1 shows
the distribution of $q$ values near the end of a simulation with and without
hierarchy.

Even in a strongly hierarchical society, revolutions can happen, and then a 
new hierarchy builds up. Thus averaging for one individual over very long time
would give an average winning probability $q$ near 1/2, for all people. The
proper way to distinguish between hierarchical and non-hierarchical societies
is thus the above snapshot method, where the inequality $\sigma$ is determined
from all $q$ values at one given moment, via eq.(2).

F. Schweitzer, private communication, has criticized the symmetry built into
this model, where the probability to belong to the leaders (high $q$) is the 
same as the one to belong to the followers (low $q$). A monarchy needs one king
with many subjects. This is only one of many questions which are still open for further modelling.

\centerline{Consensus:}

A democracy needs both a stable government and a viable opposition. But for 
a selection comittee trying to fill a professor position at the university 
it is much nicer if finally a consensus is established about who are the best
candidates. Thus consensus may be a good or a bad thing, depending on the
application. This section deals with some models of opinion formation where,
depending on details, a complete consensus is or is not found \cite{axelrod}.

Imagine that the possible opinions of a large set of people $i = 1,2, \dots, N$
are described by a real number $S$ between zero and unity; it politics this
correponds to the traditional left-right classification. Now these people 
talk to each other and try to convince each other of their opinions. Normally
people with vastly different opinions will not agree on any compromise but
when the opinions are simular then a compromise is possible. For example,
if two people $i$ and $k$ have similar opinions $S_i$ and $S_k$ such that 
their difference $|S_i-S_k|$ is smaller than some small limit $\epsilon$
(''bounded confidence''), then
a reasonable compromise is that both accept the average $(S_i+S_k)/2$ as their
new opinions. This rule can be generalized to several people agreeing within 
$\epsilon$ with opinion $S_i$, or to a vector of several opinion variables
per person, instead of merely one $Si$ \cite{hegselmann,deffuant}.

With simultaneous updating of 625 initially random opinions, \cite{hegselmann}
found dozens of final opinions for $\epsilon = 0.01$, two opinions 
(''polarization'') for $\epsilon = 0.15$, and one opinion (''consensus'')
for $\epsilon = 0.25$. Thus, the more tolerant people are, the higher is the 
chance for consensus, certainly a plausible result showing that the model is 
reasonable. Similar results were found by the French group \cite{deffuant},
giving about $1/(2 \epsilon)$ different final opinions in a similar model. 
In both cases, everybody could interact with everybody, like at a long 
conference reception where nobody can sit and thus everybody walks around a lot.

In the opposite extreme, people site on a lattice and talk only with their
nearest neighbours. And in between is the case of slow diffusion when after
every chat with a neighbour everybody tries to move to an empty neighbour
place, similar to \cite{schelling}. Computationally it is simpler to model
opinions as discrete integers $S = 1,2,3, \dots q$ such that only neighbouring
opinions $S \pm 1$ can influence opinion $S$. Thus geometric space is a 
square lattice, and opinion space is a one-dimensional chain. The role of the
previous parameter $\epsilon$ is now played by $1/q$.

This discrete dynamics was used in particular for the Sznajd model \cite{sznajd}
where, however, opinion $S_i$ is not influenced by its neighbours (as in Ising 
models \cite{schelling,callen}), but instead it influence them. Thus in a 
simple version,
one site $i$ forces its opinion $S_i$ onto all those nearest neighbours on 
the lattice which have the opinion $S_i \pm 1$. Information thus flows from
inside out, instead of the usual information flow 
from outside inwards. The people
are now more like missionaries trying to convince others, not negotiators 
looking for a compromise or opportunists accepting the majority opinion of
their neighbours (voter models).

Most of the research on Sznajd models uses, however, the original principle of
''united we stand, divided we fall''. Then two neighbours with the same opinion
convince their six neighbours on the square lattice. If the two people in 
the middle have divided opinions, they do not convince anybody. Similarly,
the children obey perhaps their parents of both parents agree with each other;
if mother says something different from father, the children are more free to do
what they want. This model, which has also been simulated in one and three
dimensions as well as on the triangular lattice, always leads to a consensus 
when $q=2$, i.e. when all opinions are similar to each other in the sense of the
above $S \pm 1$ rule. (If we allow $q > 2$ and relax the bounded confidence
restriction $S \pm 1$ by letting a pair convince all neighbours irrepective
of the difference in opinion, then again always a consensus is found.) Since
the Sznajd model was reviewed in \cite{jasss} we now summarize mainly the more 
recent results.

The difference between simultaneous and random sequential updating of opinions
\cite{hegselmann,deffuant} is quite crucial for the Sznajd model \cite{jms}.
If simultaneously two pairs of neighbours tell me that I should vote the way
they want, and these two pairs have different opinions, then I am frustrated 
and do not change my opinion. This frustration effect makes a consensus very
difficult, one needs a very large initial majority for one opinion to find a
consensus for this opinion ($q=2$, square lattice). Thus formal committee
meetings have less chance of success than informal encounters spread over a longer
time interval. 

What if the Sznajd agents diffuse slowly on a half-filled 
square lattice under bounded
confidence, i.e. $S$ convinces only $S \pm 1$,  with $q > 2$ opinions. For $q = 
5$, usually most people at the end will have adopted the centrist opinion 3,
some the extremist opinions 1 and 5, but opinions 2 and 4 have died out 
completely. More interesting are $q=4$ opinions: One of the more centrist
opinions, like 2, wins over nearly everybody at the end, after having eaten up 
all neighbouring opinions 1 and 3. A small opposition of opinion 4 usually 
remains left.
Thus if you want to stay with the winner, observe the evolution of votes: The
one who is leading half-way through the race will get most votes at the end,
the one who is on second place at half-time will get nothing, while the third-
ranked opinion has a small set of followers at the end. The fourth-ranked 
opinion dies out soon. Fig.2 shows two typical examples on a $101 \times 101$
square lattice, with 4 and 5 opinions. (For clarity the opinions in the $q=4$
case are plotted not at 1, 2, 3, 4 but at 1.5, 2.5, 3.5 and 4.5; the 39 votes 
at 4.5 are barely visible on this scale.)

\begin{figure}[hbt]
\begin{center}
\includegraphics[angle=-90,scale=0.5]{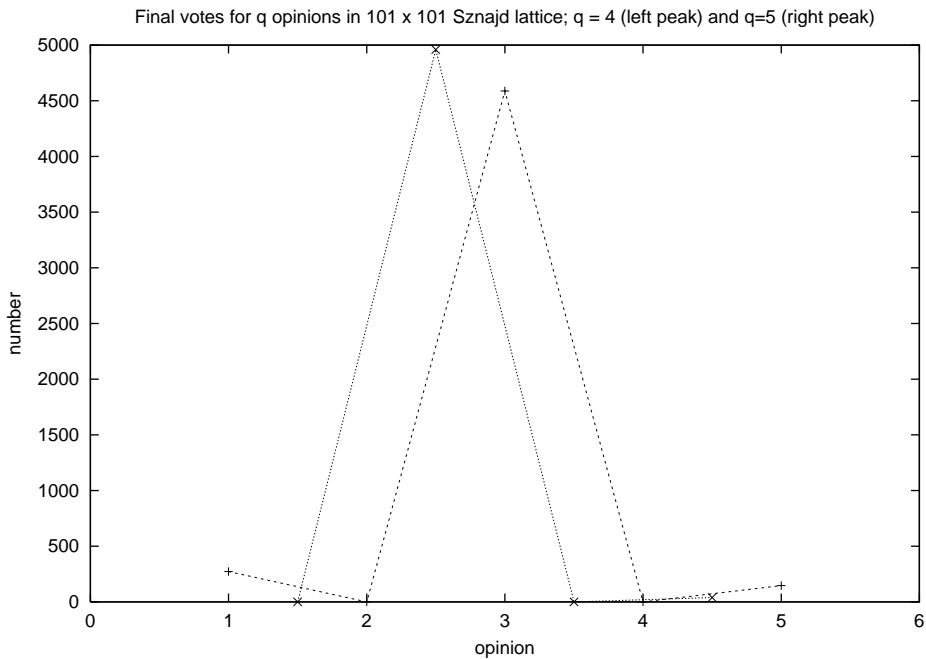}
\end{center}
\caption{Final distribution of votes for $q=4$ (solid line, x) and $q=5$ (dashed
line, +)
on a half-filled $101 \times 101$ Sznajd lattice. For $q=4$ the 
opinions are shifted to the right by 1/2 for clarity.
}
\end{figure}

Thus far opinions were exchanged only among neighbours. There are, however,
also mass media which try to influence us through advertising. This can be 
simulated \cite{schulze1} by assuming for $q=2$ that
at every iteration after the exchange with neighbours every person is flipped
into opinion 1, if opinion 1 is the one which advertises. Then for large 
lattices already a small amount of advertising is sufficient to convince the 
whole square lattice, if initially the two opinions were shared equally often. 

Physicists like to have a Hamiltonian (usually meaning just energy) for their 
models; for Sznajd this was achieved only in one dimension \cite{hamilt}.
Physicists also are accustomed that long range forces proportional to 
1/distance$^z$ facilitate phase transitions. The Sznajd model with interactions 
only between nearest neighbours on the square lattice has such a phase 
transition for $q=2$: That opinion which initially has a tiny majority,
at the end gets all votes. The introduction of long-range forces \cite{schulze2}
does not change this; in fact, $z = 1$ seems to make the transition less
pronounced than $z=4$.

\centerline{Summary:}

In the above examples, the whole human being is reduced to a simple number,
which represents her opinion (consensus models) or his history (hierarchies).
This is, of course, a great simplification, and cognitive scientists may
dislike it. Actually, more complicated models of human behaviour have been
simulated extensively as neural networks, and one could apply these neural
network techniques onto the above models: Does a person recognize before
a fight the enemy through associative memory ? How much external influence
is needed to move from one intrinsic fixed point of the neural network to 
another and thus to change opinion (consensus models). The above models ignore
these details just as Kepler's laws how the Earth circles the Sun ignore the
whole structure of the Earth. Clearly, the Earth is not point-like, Kepler
knew that, and geographers endanger their employment if they treat the Earth
as a point mass. But for the purpose of describing celestian motions, the
model of a point-like Earth is good and lead to the development of theoretical
mechanics by Newton and others later.

Humans, in contrast to the Earth, should have intelligence and thus the above
analogy with the Earth may be inappropriate. We make our own decision whether 
we smoke, drink beer, or go on a diet; and all these decisions may influence
our health and age of death. Clearly, we do not make these decisions randomly.
Nevertheless, averaged over a large population, experts have constructed life 
tables, which entered into health and life insurance business and seem to work
reasonably even though they assume humans as dying randomly. 
The first life tables were constructed centuries ago by Halley, for whom a 
famous "comet" is named.  In a similar
sense, the above computer simulations may give us information on averages over
many people, but not the fate of one specific person. For example, the Sznajd
model was used to simulate the distribution of votes among candidates in 
Brazilian elections in general, but could not predict how many votes one 
named candidate in one specific election got.
  
I hope this small selection of examples encourages the readers to enter this 
field and to invent their own simulation models. 
Thanks are due to A. Maksymowicz and K. Ku{\l}alkowski of AGH in Krak\'ow, 
Poland for hospitality when this review was drafted there.

\end{document}